\documentclass{ws-ijgmmp}

\newcommand{\be}{\begin{equation}}
\newcommand{\ee}{\end{equation}}
\newcommand{\ben}{\begin{eqnarray}}
\newcommand{\een}{\end{eqnarray}}
\newcommand{\bes}{\begin{subequations}}
\newcommand{\ees}{\end{subequations}}
\newcommand{\bn}{\begin{eqnarray}}
\newcommand{\en}{\end{eqnarray}}

\begin{document}

\markboth{Santos, and Moraes}
{Cosmology from a running vacuum model driven by a scalar field}

%
\catchline{}{}{}{}{}
%

\title{Cosmology from a running vacuum model driven by a scalar field}

\author{J. R. L. Santos}

\address{UFCG - Universidade Federal de Campina Grande - Unidade Acad\^emica de F\'isica\\
58429 – 900, Campina Grande, PB, Brazil\\
joaorafael@df.ufcg.edu.br}

\author{P. H. R. S. Moraes}

\address{ITA - Instituto Tecnol\'ogico de Aeron\'autica - Departamento de F\'isica\\ 
2228-900, S\~ao Jos\'e dos Campos, SP\\
Università degli Studi di Napoli Federico II (UNINA) - Department of Physics \\
80126, Napoli, Italy\\ 
moraes.phrs@gmail.com}

\maketitle

\begin{history}
\received{(8 April 2019)}
\accepted{(22 November 2019)}
\end{history}

\begin{abstract}
The possibility that the vacuum energy density $\rho_\Lambda$ is, indeed, varying in time, has been investigated lately from the running vacuum models perspective. Motivated by such models, in the present work, we relate the decaying vacuum energy $\rho_\Lambda(t)$ with a scalar field $\phi$. We derive the equations of motion from such a premise and implement the first-order formalism in order to obtain analytical solutions to the cosmological parameters. We show that those are in agreement with recent Planck observational data. We discuss the physical consequences of having $\rho_\Lambda(t)$ related to $\phi$.
\end{abstract}

\bigskip
Pacs: 95.30.Sf, 03.50.-z, 98.80.-k, 11.27.+d



\section{Introduction}\label{sec:int}

At current time, the Universe appears to be dark, due to the presence of dark energy and dark matter as main contributions to the Einstein's energy momentum tensor. The Planck collaboration has shown, in fact, that the $69\%$ of the Universe is filled by dark energy responsible for the observed acceleration. As it is well-known, the best model which fits the cosmological parameters derived by Planck and by other collaborations such as Dark Energy Survey \cite{des_18}, is the $\Lambda$CDM model, which is based on the cosmological constant $\Lambda$, as the source of acceleration, and also contains cold dark matter. 

Despite its success, $\Lambda$CDM model has at least two unpleasant issues. The first is known as the cosmological constant problem, which relies on the huge difference between the cosmological constant value derived from cosmological observations and from quantum field theory \cite{weinberg/1989,padmanabhan/2003}. The other issue is denominated coincidence problem, which is the fact that nowadays the amounts of dark energy and dark matter in the Universe are of the same order \cite{zlatev/1999}-\cite{pavon/2005}. A candidate to alleviate  the coincidence problem is a scalar field, in such a way that its time evolution could lead to an accelerated phase for the Universe without the need of a cosmological constant. One example of such an approach was investigated by Luongo and Muccino \cite{luongo_18}, who revised the cosmological standard model taking a scalar field Lagrangian which accounts for baryons and cold dark matter. 

Another effort to amend these problems is based on a varying cosmological ``constant'' $\Lambda = \Lambda(t)$, which is also denominated decaying vacuum. Among the several studies on decaying or running vacuum models, we highlight the contributions from Coleman et al. \cite{coleman/1980},  Rajantie et al. \cite{rajantie/2017} and Polyakov \cite{polyakov/2010}. In their seminal paper \cite{coleman/1980}, Coleman et al. investigated how vacuum decay can affect gravitation. Gravitation effects on bounces in the Standard Model of particles were later studied by Rajantie et al. in Reference \cite{rajantie/2017}. Polyakov has unveiled how vacuum is able to produce stimulated radiation in a de-Sitter space during its decaying process \cite{polyakov/2010}. 


Running vacuum models are, therefore, quite helpful in cosmology particularly because they can explain the cosmological constant problem. As it was shown in \cite{lima/2013}, the small current value of the vacuum energy density can be conceived as a result of the massive disintegration of vacuum into matter during the primordial stages of the Universe evolution.

Also, the $\Lambda$CDM model hardly provides by itself a definite explanation for the complete cosmic evolution, which involves an early and a late accelerated regimes of expansion separated by billions of years. The possibility of describing the whole cosmic history uniquely can also be attained in running vacuum models \cite{lima/2013}-\cite{lima_94} (the present authors have shown that this is also possible via $f(R,T^\phi)$ gravity, with $R$ being the Ricci scalar and $T^\phi$ the trace of the energy-momentum tensor of a scalar field \cite{ms/2016}).

As mentioned above, the $\Lambda$CDM model is also plagued by the coincidence problem. Remarkably, running vacuum models can also alleviate this issue, since according to them, the density parameters of vacuum and matter are equal in two different epochs of the Universe evolution \cite{zilioti/2018}. 

In fact, in \cite{zilioti/2018}, it was shown that running vacuum models are also a good alternative to treat the Big-Bang singularity, horizon problem, baryogenesis problem and de-Sitter instability.


It is important to remark that there are several alternative routes to describe the cosmological constant effects, such as the holographic dark energy \cite{hsu_2004}, loop quantum cosmology \cite{Oikonomou_2018, Oikonomou_2018_02}, modified gravity theories like $f(R)$ gravity \cite{Astashenok}-\cite{ODINTSOV_14}, $f(G)$ gravity \cite{ODINTSOV}, with $G$ being the Gauss-Bonnet scalar, $f(R,T)$ gravity \cite{harko} and different forms of cosmological fluids \cite{dymnikova_2001, Capozziello_2018}. The consideration of a running vacuum model in cosmology is, anyway, remarkable as one can also see \cite{freese/1987,alcaniz/2012}. Anyhow, it is worth mentioning that the origin of $\Lambda$ remains an open question with several approaches but none definitive answer \cite{abdalla}. 

In the present article, we propose a scenario with a time-dependent cosmological constant which is able to connect different phases of the Universe evolution. Our methodology is particularly based on a running cosmological constant driven by a scalar field.  

In order to introduce our ideas, let us go back to the energy conservation violation for a running vacuum model \cite{lima/2013} 

\begin{equation}\label{i1}
\dot{\rho}_m+3H\rho_m(1+\omega_m)=-\dot{\rho}_\Lambda,
\end{equation}
for $\rho_m$ being the usual matter-energy density, $H=\dot{a}/a$ the Hubble factor, $a$ the scale factor, $\omega_m$ the parameter of the equation of state (EoS) of matter, $\rho_{\Lambda}$ the vacuum energy density and dots represent time derivatives. The presence of the term $\dot{\rho}_\Lambda\neq0$ requires some energy exchange between matter and vacuum. Also, when $\rho_\Lambda$ is a constant in Equation \eqref{i1}, the standard cosmology prediction is trivially retrieved.


 Equation \eqref{i1} is clearly a consequence of the Bianchi identities applied to Einstein's field equations in the presence of a varying cosmological constant. It could also, naturally, appear from the consideration of a term $\Lambda(t)g_{\mu\nu}$, with $g_{\mu\nu}$ being the metric, to be added to the usual matter energy-momentum tensor. Such a variable $\Lambda$ was introduced in References \cite{ozer/1986,ozer/1987} with the purpose of solving the entropy problem without need for introducing some specific fields or irreversible processes. In fact, while a constant $\Lambda$ yields $Td\mathcal{S}=0$, for all the parts of the Universe in equilibrium at temperature $T$, with $\mathcal{S}$ being the entropy, a varying $\Lambda$ yields $Td\mathcal{S}\sim-d\Lambda$.

It is important to remark that such an extra contribution to the usual matter energy-momentum tensor may also appear as a result of renormalization. Take for instance Ref.\cite{linde/1979}, in which after renormalization the energy-momentum tensor $\langle\theta_{\mu\nu}\rangle$ reads $\langle\theta_{\mu\nu}\rangle=\langle T_{\mu\nu}\rangle+\epsilon(\sigma)g_{\mu\nu}$. This may be interpreted as if $\langle T_{\mu\nu}\rangle$ is the energy-momentum tensor of observable matter and $\epsilon(\sigma)g_{\mu\nu}$ is the energy-momentum tensor of the non-observable vacuum. A similar interpretation can be given to the extra terms in the effective energy-momentum tensor of the $f(R,T)$ gravity \cite{harko}.


In Eq.(\ref{i1}), the matter-energy density $\rho_m$ and the pressure $p_m=\omega_m\rho_m$ are the non-null components of the energy-momentum tensor of a perfect fluid. If we assume a scalar field $\phi$ with Lagrangian density 

\begin{equation}\label{i2}
\mathcal{L}=\frac{\dot{\phi}^{2}}{2}-V(\phi)\,,
\end{equation}
to be responsible for the matter field, as in quintessence models \cite{zimdahl/2001,ms/2014}, with $V(\phi)$ being the scalar field potential, but keep the vacuum decaying term, novel features for the Universe dynamics are expected. One can interpret it as the vacuum energy supplying the scalar field, or vice versa. As it was previously mentioned,  the development of such a formalism and the derivation of its cosmological consequences are the main goal of the present analysis. Besides, it is relevant to point that all the approaches of this investigation are based on analytical calculations. 

The structure of the paper is as follows. In Section \ref{sec:quint}, we introduce our running vacuum model and  implement the first-order formalism in order to find analytical cosmological models. After that, in Section \ref{sec:ci}  we explore the cosmological interpretation of an analytical running vacuum model. Section \ref{psp} unveils the behavior of the parameters from the power spectrum perturbations. Finally, in Section \ref{sec:con} we present our conclusions and future perspectives for the methodology here introduced.  

\section{The running vacuum model}\label{sec:quint} 

Departing from standard $\Lambda$CDM model of cosmology \cite{hinshaw/2013}, in running vacuum models, $\rho_\Lambda$ varies in time. For the evolution of $\rho_\Lambda$ it has been proposed an even power series of the Hubble rate as \cite{lima/2013}

\begin{equation}\label{qn1}
\Lambda(H)=c_0+c_2H^{2}+c_4H^{4}+...,
\end{equation}
with $c_0$ representing the dominant term when $H\sim H_0$, $H^{2},H^{4}...$ are small corrections to the dominant term, which provide a time-evolving behavior to the vacuum energy density and $c_2,c_4...$ are constants. These corrections account for quantum field theories contributions to gravity, which keep the covariance of Einstein's equations \cite{shapiro}.


 One may wonder how to overcome degeneracy between a model based on Eq.\eqref{qn1} and standard dark energy models. One possible path is the so-called cosmography method. This method uses cosmographic techniques to establish cosmological constraints on the observable Universe \cite{aviles/2014,capozziello/2019}. Moreover, it can be used to test generalized versions of General Relativity \cite{capozziello/2019}. A relevant point about cosmography is that it can distinguish among models that are compatible with experimental data. Once \eqref{qn1} will lead us to modified Friedmann equations, these can be combined to map the cosmographic parameters, by following the procedure adopted in \cite{capozziello/2019}.


We can write the vacuum energy density in terms of \eqref{qn1} as
\begin{equation}\label{qn2}
\rho_\Lambda=\frac{\Lambda(H)}{8\pi\,G}=\frac{\Lambda(H)}{2},
\end{equation}
with the second equality valid if we are working with $c=4\,\pi\,G=1$ units. 

Since here we wish to establish that the scalar field is related to the vacuum energy density, we can naturally write the action as

\be \label{qn3}
S=\frac{1}{2}\int\,d\,x^{\,4}\,\sqrt{-g}\,\left(-\frac{R}{2}-\Lambda+{\cal L}\right)\,, \qquad 
\Lambda=\Lambda(\phi)\,, \qquad {\cal L}={\cal L}\,\left(\partial_\mu\,\phi, \phi\right)\,,
\ee
which can be minimized with respect to the metric, leading to the Einstein's field equations
\be
R_{\,\mu\,\nu}-\frac{1}{2}\,g_{\,\mu\,\nu}\,R=2\,\widetilde{T}_{\,\mu\,\nu}\,,
\ee
where $R_{\mu\nu}$ is the Ricci tensor and
\be
\widetilde{T}_{\,\mu\,\nu}=T_{\,\mu\,\nu}+g_{\,\mu\,\nu}\,\rho_\Lambda\,,
\ee
with $T_{\,\mu\,\nu}$ being the energy-momentum tensor of the scalar field, such that

\begin{equation}\label{qn4}
T_{\mu\nu}=2\frac{\partial\mathcal{L}}{\partial g^{\mu\nu}}-g_{\mu\nu}\mathcal{L}\,.
\end{equation}
Here, $T_{\mu\nu}=\texttt{diag}(\rho,-p,-p,-p)$, with $\rho$ and $p$ being the matter-energy density and pressure of the other components of the Universe, as radiation and matter.

Considering that this background field depends only on time and that the Lagrangian ${\cal L}$ has the form presented in $(\ref{i2})$, we can use $(\ref{qn2})$ and $(\ref{qn4})$ to determine the density and the pressure due to the field $\phi$ as 
\be \label{qn5}
\rho=\frac{\dot{\phi}^{2}}{2}+V(\phi),
\ee
\be \label{qn6}
p=\frac{\dot{\phi}^{2}}{2}-V(\phi).
\ee

The substitution of Eqs.(\ref{qn5}) and (\ref{qn6}) in (\ref{i1}) yields

\begin{equation}\label{qn7}
\ddot{\phi}+3H\dot{\phi}+V^{\,\prime}=-\frac{\dot{\rho}_\Lambda}{\dot{\phi}},
\end{equation}
where primes denote derivations with respect to $\phi$.

Moreover, the minimization of the action \eqref{qn3} with respect to the field leads the equation of motion

\be \label{qn8}
\ddot{\phi}+3\,H\,\dot{\phi}+V^{\,\prime}=-\rho_{\Lambda}^{\,\prime}\,,
\ee
for the Friedmann-Lema\^itre-Robertson-Walker metric with null curvature \cite{hinshaw/2013}. In this last equation we made use of Eq.$(\ref{qn2})$.

From Eqs.(\ref{qn7})-(\ref{qn8}), one obtains that

\begin{equation}\label{qnex}
\frac{\dot{\rho}_\Lambda}{\dot{\phi}}=\rho_{\Lambda}^{\,\prime},
\end{equation}
which can indeed be verified by recalling that $\rho_\Lambda=\rho_\Lambda(t)$.

Following the procedures adopted in \cite{lima/2013}, the Friedmann equations for such a configuration are
\be \label{qn9}
\rho+\rho_\Lambda=\frac{3}{2}\,H^{\,2}\,, 
\ee
\be \label{qn10}
p-\rho_\Lambda=-\dot{H}-\frac{3}{2}\,H^{\,2}\,.
\ee 
Consequently, by substituting $(\ref{qn5})$ and $(\ref{qn6})$ into $(\ref{qn9})$ and $(\ref{qn10})$ we determine that
\be \label{qn11}
V=\frac{1}{2}\,(3H^{\,2}-\dot{\phi}^{\,2})-\rho_\Lambda\,,
\ee
\be \label{qn12}
\dot{H}=-\dot{\phi}^{\,2}\,.
\ee

\subsection{First-order formalism implementation}\label{ss:fofi}

In order to obtain the solutions of the present model, we will implement the first-order formalism. It was shown in \cite{bazeia/2006} how to determine first-order differential equations involving one scalar field, whose solutions satisfy the equations of motion. It can be seen in the literature that an advantage of this method is the attainment of analytical cosmological parameters \cite{ms/2014}.

So, in order to implement the first-order formalism in the context of running vacuum models, let us consider that the Hubble parameter can be written as \cite{bazeia/2006}
\be \label{qn13}
H=-W(\phi)\,,
\ee
where $W$ is a generalized function of the field $\phi$, also known as superpotential. Then, Eq.$(\ref{qn12})$ yields the first-order differential equation

\be \label{qn14}
\dot{\phi}=W'.
\ee

The previous definition for $H$ results in the potential
\be \label{qn15}
V=\frac{1}{2}[W^{\,2}-W'^{\,2}-\,\left(c_0+c_2\,W^{\,2}+c_4\,W^{\,4}\right)]\,,
\ee
for $\rho_{\Lambda}$ up to $c_4$ in Eqs.(\ref{qn1})-(\ref{qn2}). We can verify that Eqs.$(\ref{qn13})$-$(\ref{qn15})$ satisfy $(\ref{qn7})$-$(\ref{qn8})$. Therefore, the solutions of the first-order equation $(\ref{qn14})$ automatically obey the equation of motion for the field $\phi(t)$. 

Moreover, the EoS parameter $\omega=p/\rho$ and the acceleration parameter for this model are
\be \label{qn16}
\omega=-\left[\frac{2\,W'^{\,2}}{c_4\,W^{\,4}+(c_2-3)\,W^{\,2}+c_0}+1\right]\,,
\ee
\be \label{qn16_1}
q=-\frac{\dot{H}}{H^{\,2}}-1=\left(\frac{W^{\,\prime}}{W}\right)^{\,2}-1\,,
\ee
respectively.

\subsubsection{Example}

As an example to unveil the applicability of such a formalism, let us work with
\be \label{qn17}
W=b_1\,\left(\phi-\frac{\phi^{\,3}}{3}\right)+b_2\,,
\ee
where $b_1$ and $b_2$ are real constants. This form for $W$ was used in several works about field theory and cosmology, as one can see \cite{ms/2016},\cite{bazeia/2013, bazeia_04, bazeia_042}, for instance.  

The relevance of this form of $W(\phi)$ raises in the fact that it can generate the simplest example of a topological defect solution when we deal with a classical theory for static fields. As a matter of clarification, let us briefly review some points on classical field theory for defects in Minkowski space-time with $1+1$ dimensions. By considering an action with the form 
\be
S_c=\int\,dt\,d\,x\,\left[\frac{1}{2}\,\partial_{\mu}\,\phi\,\partial^{\,\mu}\phi-V_c(\phi)\right]\,; \qquad \phi=\phi(x,t)
\ee
its equation of motion is 
\be
\ddot{\phi}-\phi_{\,x\,x}+V_c^{\,\prime}=0\,.
\ee
So, considering a static field, the previous equation is reduced to
\be
\phi_{\,x\,x}=V_c^{\,\prime}\,; \qquad \phi_{\,x}=\frac{d\,\phi}{d\,x}\,,
\ee
which can be integrated once yielding
\be
\phi_{\,x}=\pm\,W^{\,\prime}\,; \qquad V_c\equiv \frac{W^{\,\prime\,2}}{2}\,.
\ee
Then, if we work with $W(\phi)$ from Eq. $(\ref{qn17})$, the last expressions are rewritten as
\be
\phi_{\,x}=\pm\,b_1\,\left(1-\phi^{\,2}\right)\,; \qquad V_c=b_1^{\,2}\frac{\left(1-\phi^{\,2}\right)^{\,2}}{2}\,,
\ee
whose analytical solutions are
\be
\phi(x)=\pm\,\tanh(b_1\,x+b_3)\,.
\ee
These classical solutions are known as topological defects or a pair kink/anti-kink, and the mass of the defect can be calculated from the potential $V_c$, as follows
\be
m_{\phi}^{\,2}\equiv V_c^{\,\prime\,\prime}(\phi=\pm1)=4\,b_1^{\,2}\,.
\ee

\begin{figure}[ht!]
 \centering
 \includegraphics[width=0.5\columnwidth]{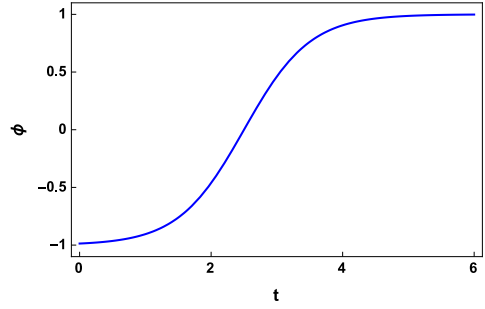}
 \caption{Field $\phi$ for $b_1= 1$ and $b_3=-2.5$. We can see that the scalar field smoothly variates from one asymptotical value to another during its time evolution.  }
 \label{Fig0}
\end{figure}

Moving back to our procedure to find cosmological solutions, we can substitute Eq.$(\ref{qn17})$ into the first-order differential equation $(\ref{qn14})$, leading to
\be \label{qn18}
\dot{\phi}=b_1\,\left(1-\phi^{\,2}\right)\,,
\ee
whose analytical solution is
\be \label{qn19}
\phi(t)=\tanh\,(b_1\,t+b_3)\,,
\ee
with constant $b_3$. The features of $\phi$ (or inflaton field) are presented in Fig.\ref{Fig0} where we see this cosmological solution has a kink-like profile. The scalar inflaton represents a homogeneous background responsible for the evolution of the cosmological parameters. As we have two possible asymptotic values for this inflaton solution, then we can build a Universe with two different inflationary regimes, one for small values of time and another one for late time. As we are going to see, the smooth transition between these inflationary regimes makes the Universe to continuously pass through other stages of its evolution, such as the radiation and matter eras.

Now that we have $W$ and $\phi$ in hands we are able to find
 
\be \label{qn20}
H=\frac{b_1}{3} \, \tanh (b_1 t+b_3) \left[\tanh ^2(b_1 t+b_3)-3\right]-b_2\,.
\ee
Moreover, 

\bn \label{qn21}
a(t) &=&\frac{\exp\, \left[\frac{1}{6}\text{sech}^2\,(b_1 t+b_3)-\, b_2\, t\right]}{\,\cosh^{\,4}\,(b_1 t+b_2)}\,.
\en

Furthermore, we find
\ben\label{omega}
&&\omega = -1-2\,[b_1 \text{sech}^2\,(b_1\, t+b_3)]^2\\ \nonumber
&&\times\bigg\{(c_2-3) \left\{b_2-b_1\left[\frac{1}{3}\, \tanh ^3(b_1 t+b_3)+\, \tanh (b_1 t+b_3)\right]\right\}^2 \\ \nonumber
&&+c_4 \left\{b_2-b_1\left[\frac{1}{3}\, \tanh ^3(b_1 t+b_3)+\, \tanh (b_1 t+b_3)\right]\right\}^4+c_0\bigg\}^{\,-1}\, \nonumber
\een
and
\be \label{acc}
q=\frac{9\,b_1^2 \,\text{sech}^4(b_1\, t+b_3)}{\left(3\,b_2-b_1\, \tanh ^3(b_1 t+b_3)+3\,b_1\, \tanh (b_1\, t+b_3)\right)^2}-1\,,
\ee

\noindent as the analytical EoS and the acceleration parameters, respectively.

\begin{figure}[ht!]
 \centering
 \includegraphics[width=0.5\columnwidth]{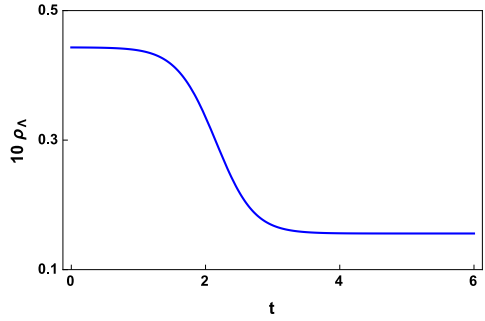}
 \caption{ Time evolution of the vacuum energy density for $b_1= 1$, $b_2=-1$, $b_3=-2.5$, $c_0=0.03$, $c_2=0.01$ and $c_4=0.004$.}
 \label{Fig4}
\end{figure}

Moreover, Equation $(\ref{qn20})$ and Equation $(\ref{qn2})$ yield the following relation for the vacuum energy density:

\ben
&&\rho_{\Lambda}=\frac{c_2}{2} \left\{b_1 \left[\tanh (b_1 t+b_3)-\frac{1}{3} \tanh ^3(b_1 t+b_3)\right]+b_2\right\}^2\\ \nonumber
&&+\frac{c_4}{2}\, \left\{b_1 \left[\tanh (b_1 t+b_3)-\frac{1}{3} \tanh ^3(b_1 t+b_3)\right]+b_2\right\}^4+\frac{c_0}{2}\,,
\een
whose evolution in time can be seen in Figure \ref{Fig4}. 

\begin{figure}[h!]
\centering
\includegraphics[width=0.5\columnwidth]{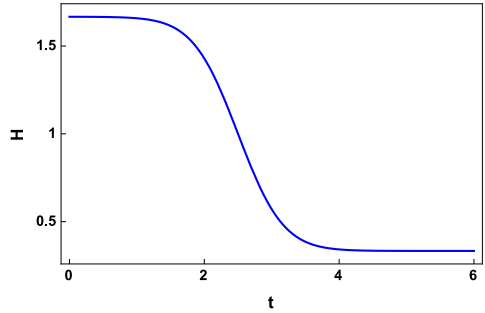}
\caption{Time evolution of the Hubble parameter for $b_1= 1$, $b_2=-1$ and $b_3=-2.5$.}
\label{Fig1}
\end{figure}

\begin{figure}[h!]
\centering
\includegraphics[width=0.5\columnwidth]{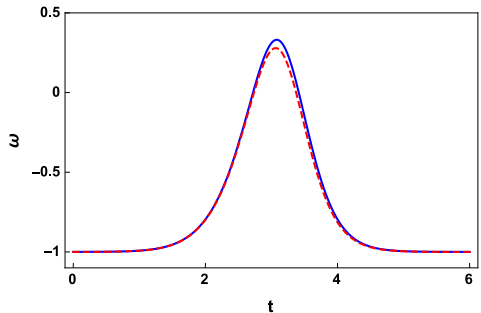}
\caption{Time evolution of the EoS parameter for $b_1= 1$, $b_2=-1$, $b_3=-2.5$, $c_0=0.03$, $c_2=0.01$ and $c_4=0.004$ (blue solid curve) and for $b_1= 1$, $b_2=-1$, $b_3=-1.5$ and $\Lambda(\phi)=0$ (red dashed curve).}
\label{Fig3}
\end{figure}

 Besides that, the features of $H$ and $\omega$ can be appreciated in Figs.\ref{Fig1} and \ref{Fig3}, with the latter presenting also a curve for the case $\Lambda(\phi)=0$ (red dashed curve). We can observe that the field $\phi(t)$ determined in Eq.$(\ref{qn19})$ leads to a Hubble parameter which exhibits a kink-like profile. Such a profile means that $H$ is approximately constant for early and late cosmic times, exhibiting a behavior consistent with two accelerated expansion eras of the Universe. More details about this behavior are presented in the next section. 

\section{Cosmological interpretations}\label{sec:ci}

In the present section we will interpret our cosmological solutions obtained in the previous section and show their feasibility. 

Let us start by analyzing Figure \ref{Fig4}. It shows the time evolution of the energy density of the running cosmological ``constant''. We can see that as time passes by, the density of $\Lambda(t)$ decreases until it attains a minimum constant value. In this way, the late-time behavior of $\Lambda$ retrieves what is expected in the standard model of cosmology.

Fig.\ref{Fig1} shows the time evolution of the Hubble parameter. For small values of $t$ we see that it is approximately constant. The standard cosmology model  states that the primordial Universe passed through an inflationary era. Such a scenario  was originally proposed in \cite{guth/1981} with the purpose of solving the horizon and flatness problems. 

During the inflationary era, the dynamics of the Universe is such that it expands in an accelerated way. The scale factor in this era can be written as $a\sim e^{H_it}$ (as it can be checked in Reference \cite{ryden/2003}), with $H_i$ being the Hubble parameter value during inflation. In other words, standard cosmology states that during inflation, the Hubble parameter remains approximately constant. From Figure \ref{Fig1}, we can see that our model prediction, in this sense, is the same as in the standard model of cosmology.

According to standard cosmology, the non-accelerated stages of the Universe expansion, namely, radiation and matter dominated eras, are described by a decreasing Hubble parameter $H\sim t^{-1}$. This can be easily derived by substituting the radiation and matter parameters of the EoS of the Universe, $\omega=1/3$ and $\omega=0$, respectively, in the standard Friedmann equations \cite{ryden/2003}. Such a decreasing function of time is the behavior of our solution for $H$ in the non-accelerated stages of the Universe expansion, as one can see from Fig.\ref{Fig1}.

As time passes by, $H$ becomes constant once again. As we mentioned above, this implies an accelerated expansion. Indeed, recent data on the fluctuations of the temperature of the cosmic microwave background show that the Universe is currently passing through a stage of accelerated expansion \cite{hinshaw/2013}.

These well-behaved features are also reflected in the EoS parameter in Fig.\ref{Fig3}. According to our model, $\omega\sim-1$ during the inflationary era, which is the expected result \cite{cognola/2009,namjoo/2012}. The EoS parameter then increases until it reaches $\sim1/3$, which is the value expected for a radiation dominated universe \cite{ryden/2003}. As time passes by, $\omega$ returns to $-1$, in accordance with recent observational data \cite{hinshaw/2013}. 

It is relevant to say that up to this point there is no considerable difference between the cosmological parameters derived from $\Lambda=0$ and those with a cosmological constant driven by a scalar field. Therefore, with the purpose of unveiling the importance of the cosmological constant in the context of a vacuum decaying model, let us approach a power spectrum analysis of this system.

\section{Features of the power spectrum perturbations}
\label{psp}

In order to complete the cosmological interpretations of our model and with the purpose of remarking its importance and feasibility, we are going to analyze the physical parameters of the power spectrum perturbations. 
 
We consider that the small perturbations regime occurs in the dark energy era, so we have $\rho_{\,\Lambda}\approx \mbox{ constant}$ and the EoS parameter $\omega \approx -1$, as one can check  Figs.\ref{Fig4} and \ref{Fig3}, respectively. In such a regime, $\dot{\rho}_{\Lambda} \approx 0$ and, consequently, we assume that the standard calculations of small perturbations are still valid. 

The cosmological features of the power spectrum can be extracted from the first two slow-roll parameters, which are explicitly written as \cite{ellis}

\be \label{slr}
\epsilon=\frac{1}{4}\,\left(\frac{V^{\,\prime}}{V}\right)^2\,; \qquad \eta = \frac{1}{2}\,\frac{V^{\,\prime\,\prime}}{V}\,,
\ee
since we are working with $4\,\pi\,G=1$ units. 

The standard inflationary scenario requires that the strength of the tensor perturbations is connected with the magnitude of the energy density. Besides, the power spectrum for scalar perturbation of a one field coupling is given by \cite{ellis}
\be
P_{\zeta}=\frac{H^{\,4}}{4\,\pi^{\,2}\,\dot{\phi}^{\,2}}\,
\ee
when we talk about quantities which are determined at the horizon crossing \cite{ellis,mukhanov}. 

Moreover, another relevant parameter is the so-called scalar spectral index whose form is \cite{ellis}
\be \label{ssp0}
n_s=1-6\,\epsilon +2\,\eta.
\ee
This parameter is important as a test for cosmological models, since it is directly measured from the cosmic microwave background, as described in the last set of data from Planck collaboration, which established that $n_s= 0.9665\,\pm\, 0.0038$ \cite{planck18}. 

Another parameter related to tensor perturbations is the tensor-scalar ratio, which, for a one scalar field Lagrangian coupled with General Relativity, reads \cite{ellis}

\be \label{ssp1}
r=\frac{P_T}{P_{\,\zeta}}\,, \qquad P_T=16\,\left(\frac{H}{2\,\pi}\right)^{\,2}\,.
\ee
This parameter has also been measured by Planck collaboration, which revealed that $r < 0.09$ \cite{planck15}. Moreover, the Starobinsky $R^2$ inflationary model \cite{planck15,starobinsky} predicts values for $r$ covering the range $r \in [0.003, 0.005]$.

By setting the constants of our model as $b_1= 2.1$, $b_2=-2.8$, $b_3=-3.3$, $c_0=5.6$, $c_2=0.0026$, $c_4=0.0012$ and substituting the Hubble parameter $H$, the field $\phi$ and the potential $V$ into ($\ref{omega}$), ($\ref{acc}$), ($\ref{slr}$), ($\ref{ssp0}$) and  ($\ref{ssp1}$), yields
\be\label{eqimp}
n_s\approx0.9650\,, \qquad r\approx0.0035\,, \qquad \omega\approx-0.9962\,,  \qquad q_0 \approx -0.9998\,,
\ee
for $t=3$. 

However, if we consider the same analysis for $c_0=c_2=c_4=0$, we find the following
\be
n_s \approx 1.0598 \,, \qquad r \approx 0.0035\,, \qquad \omega \approx -0.9999\,, \qquad q_0 \approx -0.9998\,.
\ee

The last set of parameters reveals that the scalar spectral index is out of the experimental range covered by Planck collaboration. Besides, in 2015, the Planck collaboration concluded that polynomial scalar field potentials with powers $>2$ are not compatible with these previous cosmological parameters \cite{planck15}. Therefore, we clearly see that the constants $c_0$, $c_2$ and $c_4$ were essential for restoring a well behaved physical scenario. The values for $\omega$ and $q_0$ from Eq.($\ref{eqimp}$) are also compatible with parameters derived from cosmography constraints, as one can see in Tables IV and V of \cite{aviles/2012} and in Fit (3) of Table II of \cite{gruber}.  

These calculations unveil the relevance of a running $\Lambda$ driven by a scalar field as a possible description for the current accelerated phase of the Universe expansion. It is interesting how such a mechanism can also be used to rescue the one scalar field inflation, corroborating with the approach presented in \cite{almeida/2017}, where the authors studied how a Lorentz-breaking parameter term in the dynamic of the scalar field Lagrangian may lead to a scalar spectral index as well as to a tensor-scalar ratio compatible with the recent Planck data.

\section{Discussion}\label{sec:con}

In the present article, we have applied the first-order formalism for a running vacuum model. The cosmological outcomes are well behaved and in accordance with observations. They indicate a complete cosmological history for the Universe evolution.

We have  invoked a generalized function $W( \phi)$,  referred to as superpotential, whose correspondent first-order differential equation enables us to obtain a proper form for the Hubble parameter $H(t)$. This parameter allowed us to obtain the analytical forms for the scale factor $a(t)$ and EoS parameter $\omega(t)$. The vacuum energy density $\rho_\Lambda(t)$ was also derived in an analytical form and Figure \ref{Fig4} maps its evolution with respect to time. 

Models with varying ``constants'' have shown to yield interesting and testable results, as follows. High-quality absorption lines seen in the spectra of distant quasars may allow one to probe time variations of fundamental constants. In \cite{chand/2004}, the authors presented the results from a detailed many-multiplet analysis of $18$ quasars, in the redshift range $0.4\leq z\leq2.3$, to detect the possible variation of the fine-structure constant $\alpha$. They found, as a strong constraint, that $\Delta\alpha/\alpha\sim(-0.06\pm0.06)10^{-5}$. The observations related to the variation of fundamental constants were used to impose constraints on $f(\mathcal{T})$ gravity models, with $\mathcal{T}$ being the torsion scalar, in \cite{nunes/2017}. 

Moreover, in \cite{mota/2004}, the authors have studied the space-time evolution of the fine structure constant inside evolving spherical overdensities in a $\Lambda$CDM Friedmann-Lema\^itre-Robertson-Walker Universe using the spherical infall model. The variation of the ratio of the proton mass to the electron mass and the strong coupling, fine structure and Newtonian gravitational constants was computed within the context of varying cosmological constant models in Reference \cite{fritzsch/2017}. For a review on the subject of varying constants and its relation to gravity and cosmology, we recommend Reference \cite{uzan/2011}.

Models with variation of the vacuum energy have become popular recently. In \cite{szydlowski/2015}, a particular form for $\Lambda(t)$ was constructed from quantum mechanical principles and some cosmological parameters were derived and confronted with observational data. The evolution of matter density perturbations and the cosmic star formation rate for $\Lambda(t)\sim H^2$ were calculated in \cite{bessada/2013}. In Reference \cite{szydlowski/2017}, a model with both decaying vacuum energy and dark matter was proposed. Besides, there are also studies concerning the decaying of the cosmological constant with the cosmic microwave background temperature \cite{Sonoda}.


It is important to remark that in these models, the assumption $\dot{\rho}_\Lambda\neq0$ requires necessarily some sort of energy exchange between matter and vacuum or vice-versa. That is to say that relativistic and non-relativistic matter are created in the Universe as a consequence of vacuum decaying.


Here we have considered a scalar field to be related to decaying vacuum energy. Eq.\eqref{qnex}, as far as the authors know, is a novelty in the literature. It relates the time derivatives of $\rho_\Lambda$ and $\phi$ and since those physical quantities are strongly related, the ratio between them can be written merely as $\rho_\Lambda'$. A relation between the decaying vacuum energy density and the scalar field was the main premise here.

The consequences of such a strong connection are remarkable. A cosmological model in agreement with theoretical predictions and cosmological observations was obtained. Its features are connected to the whole history of the Universe evolution, in the sense that they are able to describe inflation, radiation, matter and dark energy eras, in an analytical and continuous form. Such an attainment is impossible via standard gravity.

In order to obtain (\ref{eqimp}) we had to fix a value for the time and we have chosen $t=3$. It is worth to remark that this choice was not arbitrary. Let us carefully discuss this question in the following.

From the Planck satellite observations \cite{planck18}, the present value of the EoS parameter is  $-1.019^{+0.075}_{-0.080}$. From this result, together with Eq.\eqref{omega}, when $t=2.674$, the EoS parameter enters a region of acceptable values. Since the Planck data we are working with refers to the present values of the concerned parameters, we can say that in our graphics and equations, the present time is somewhere after $t=2.674$. For the sake of simplicity and also taking into account that the acceleration of the Universe expansion - that according to standard Friedmann equations, occurs when $\omega<-1/3$ - started some billion years ago, we took $t=3$ in order to calculate the numerical values of the parameters.

The results are remarkable. The numerical values obtained for $n_s$ and $r$ are in accordance with Planck observational data. Our value for $r$ is also within the range of values theoretically predicted from the $R^2$ gravity. Furthermore, the present value for $\omega$ is in accordance with the observed cosmic acceleration. As far as we know, our approach is a new route to rescue the one scalar field inflationary models, complementing the beautiful work by Ellis et al. \cite{ellis}.  We also point that an investigation about the evolution of $\dot{\phi}$ for modified theories of gravity would be another interesting subject for future works, enabling one to characterize different phases of the Universe depending on its rolling behavior. Such an approach could lead to analytical scenarios which evade the slow-roll approximation as the ones studied in \cite{motohashi, galvez}.

It is important to say that a similar discussion of these subjects was recently performed by Basilakos  et al. \cite{bms_19}, where the authors also analyze that the actual dark energy expansion is compatible with a cosmological constant driven by a scalar field. Moreover, another interesting work based on the scalar field description of $\Lambda$ was addressed by Maia and Lima \cite{ml2002}, where the authors showed how cosmology can be driven from a de Sitter vacuum state to a modified FRW metric. 

The results here presented open a new window to work with cosmological models coupled with a scalar field. We believe that such an approach can be used in the context of hybrid inflation for standard and generalized theories of gravity in future works. \\

\section*{Acknowledgments}

J. R. L. S. would like to thank CNPq for financial support, grant 420479/2018-0, and CAPES. P. H. R. S. M. would like to thank S\~ao Paulo Research Foundation (FAPESP), grant 2015/08476-0, for financial support. The authors are grateful to G. Dellarole for some fruitful discussions regarding the possibility of investigating the scalar field dynamics as related to decaying-vacuum models. The authors would also like to thank the anonymous referee for his/her useful comments.


\begin{thebibliography}{100}

\bibitem{planck18} N. Aghanim  et al. [Planck Collaboration], Planck 2018 results. VI. Cosmological parameters, astro-ph.CO/1807.06209.
\bibitem{des_18} E. Baxter et al. [Dark Energy Survey], Dark Energy Survey Year 1 Results, astro-ph.CO/1802.05257.
\bibitem{weinberg/1989} S. Weinberg, The cosmological constant problem, {\it Rev. Mod. Phys.}  {\bf 61} (1989), 1.
\bibitem{padmanabhan/2003} T. Padmanabhan, Cosmological constant - the weight of the vacuum, {\it Phys. Rep.}  {\bf 380}  (2003), 235.
\bibitem{zlatev/1999} I. Zlatev, L. Wang, and P. J. Steinhardt, Quintessence, Cosmic Coincidence, and the Cosmological Constant, {\it Phys. Rev. Lett.}  {\bf 82} (1999), 896.
\bibitem{luongo_18}O. Luongo, and M. Muccino, Speeding up the Universe using dust with pressure, {\it Phys. Rev. D}  {\bf 98} (2018), 103520.
\bibitem{chimento/2003} L. P. Chimento, A. S. Jakubi, D. Pavon, and W. Zimdahl, Interacting quintessence solution to the coincidence problem, {\it Phys. Rev. D}  {\bf 67}  (2003), 083513.
\bibitem{pavon/2005} D. Pav\'on, and W. Zimdahl, Holographic dark energy and cosmic coincidence, {\it Phys. Lett. B}  {\bf 628} (2005), 206.
\bibitem{coleman/1980} S. Coleman, and F. de Luccia, Gravitational effects on and of vacuum decay, {\it Phys. Rev. D}  {\bf 21} (1980), 3305.
\bibitem{rajantie/2017} A. Rajantie, and S. Stopyra, Standard model vacuum decay with gravity, {\it Phys. Rev. D}  {\bf 95} (2017),  025008.
\bibitem{polyakov/2010} A. M. Polyakov, Decay of vacuum energy, {\it Nucl. Phys. B}  {\bf 834} (2010), 316.
\bibitem{lima/2013} J. A. S. Lima, S. Basilakos, and J. Sola, Expansion history with decaying vacuum: a complete cosmological scenario, {\it MNRAS}  {\bf 431} (2013), 923.
\bibitem{zilioti/2018} G. J. M. Zilioti, R. C. Santos, and J. A. S. Lima, From de Sitter to de Sitter: Decaying Vacuum Models as a
Possible Solution to the Main Cosmological Problems, {\it Adv. High Ener. Phys.}  {\bf 2018} (2018), 6980486.
\bibitem{lima_94} J. A. S. Lima, and J. M. F. Maia, Deflationary cosmology with decaying vacuum energy density, {\it Phys. Rev. D}  {\bf 49} (1994),  5597.
\bibitem{ms/2016} P. H. R. S. Moraes, and J. R. L. Santos, A complete cosmological scenario from $f(R,T^{\phi})$ gravity theory, {\it Eur. Phys. J. C}  {\bf 76} (2016),  60.
\bibitem{hsu_2004} S. D. H. Hsu, Entropy bounds and dark energy, {\it Phys. Lett. B}  {\bf 594}  (2004), 13.
\bibitem{Oikonomou_2018} K. Kleidis, and V. K. Oikonomou, Loop quantum cosmology scalar field models, {\it Int. J. Geom. Meth. Mod. Phys.}  {\bf 15}  (2018), 1850071.
\bibitem{Oikonomou_2018_02} K. Kleidis, and V. K. Oikonomou, Loop quantum cosmology-corrected Gauss–Bonnet singular cosmology, {\it Int. J. Geom. Meth. Mod. Phys.}  {\bf 15} (2018), 1850064.
\bibitem{Astashenok} A. V. Astashenok, K. Mosani, S. D. Odintsov, and G. C. Samanta, Gravitational collapse in General Relativity and in $R^{\,2}$ - gravity: A comparative study, {\it Int. J. Geom. Meth. Mod. Phys.}  {\bf 03} (2019), 1950035.
\bibitem{Sebastiani} L. Sebastiani, and R. Myrzakulov, $F(R)$-gravity and inflation, {\it Int. J. Geom. Meth. Mod. Phys.}  {\bf 12} (2015), 1530003.
\bibitem{ODINTSOV_14} S. Nojiri, and S. D. Odintsov, Accelerating cosmology in modified gravity: From convenient $F(R)$ or string-inspired theory to bimetric $F(R)$ gravity, {\it Int. J. Geom. Meth. Mod. Phys.}  {\bf 02}  (2014), 1460006.
\bibitem{ODINTSOV} S. Nojiri, and S. D. Odintsov, INTRODUCTION TO MODIFIED GRAVITY AND GRAVITATIONAL ALTERNATIVE FOR DARK ENERGY, {\it Int. J. Geom. Meth. Mod. Phys.}  {\bf 04} (2007), 115.
\bibitem{harko} T. Harko, F. S. N. Lobo, S. Nojiri, and S. D. Odintsov, $f(R,T)$ gravity, {\it Phys. Rev. D}  {\bf 84} (2011), 024020.
\bibitem{dymnikova_2001} I. Dymnikova, and M. Khlopov, Decay of cosmological constant in selfconsistent inflation, {\it Eur. Phys. J. C}  {\bf 20} (2001), 139.
\bibitem{Capozziello_2018} S. Capozziello, C. A.  Mantica, and L. G. Molinari, Cosmological perfect fluids in Gauss–Bonnet gravity, {\it Int. J. Geom. Meth. Mod. Phys.}  {\bf 16} (2019), 1950008.
\bibitem{freese/1987} K. Freese, F. C. Adams, J. A. Frieman, and E. Mottola, Cosmology with decaying vacuum energy, {\it Nuc. Phys. B}  {\bf 287}  (1987),  797.
\bibitem{alcaniz/2012} J. S. Alcaniz, H. A. Borges, S. Carneiro, J. C. Fabris, C. Pigozzo, and W. Zimdahl, A cosmological concordance model with dynamical vacuum term, {\it Phys. Lett. B}  {\bf 716}  (2012), 165.
\bibitem{abdalla} R.G. Landim, and E. Abdalla, Metastable dark energy, {\it Phys. Lett. B}  {\bf 764}  (2017), 271.
\bibitem{ozer/1986} M. \"{O}zer, and M.O. Taha, A possible solution to the main cosmological problems, {\it Phys. Lett. B}  {\bf 171}  (1986), 363.
\bibitem{ozer/1987} M. \"{O}zer, and M.O. Taha, A model of the universe free of cosmological problems, {\it Nucl. Phys. B}  {\bf 287}  (1987), 776.
\bibitem{linde/1979} A. D. Linde, Phase transitions in gauge theories and cosmology, {\it Rep. Prog. Phys.}  {\bf 42}  (1979), 861.
\bibitem{zimdahl/2001} W. Zimdahl, Diego Pav\'on, and L. P. Chimento, Interacting quintessence, {\it Phys. Lett. B}  {\bf 521}  (2001), 133.
\bibitem{ms/2014} P. H. R. S. Moraes, and J. R. L. Santos, Two scalar field cosmology from coupled one-field models, {\it Phys. Rev. D}  {\bf 89}  (2014),  083516.
\bibitem{hinshaw/2013} G. Hinshaw et al., NINE-YEAR WILKINSON MICROWAVE ANISOTROPY PROBE (WMAP)
OBSERVATIONS: COSMOLOGICAL PARAMETER RESULTS, {\it Astrophys. J. Supp.} {\bf 208}  (2013), 18.
\bibitem{shapiro} I. L. Shapiro, and J. Sola, On the possible running of the cosmological ''constant", {\it Phys. Lett. B}  {\bf 682}  (2009), 105.
\bibitem{aviles/2014} A. Aviles, A. Bravetti, S. Capozziello, and O. Luongo, Precision cosmology with Padé rational approximations: Theoretical predictions versus observational limits, {\it Phys. Rev. D}  {\bf 90}  (2014),  043531.
\bibitem{capozziello/2019} S. Capozziello, R. D'Agostino, and O. Luongo, Extended gravity cosmography, {\it Int. J. Mod. Phys. D}  {\bf 28}  (2019), 1930016.
\bibitem{bazeia/2006} D. Bazeia, C. B. Gomes, L. Losano, and R. Menezes, First-order formalism and dark energy, {\it Phys. Lett. B}  {\bf 633}  (2006),  415.
\bibitem{bazeia/2013} D. Bazeia, M. A. Gonz\'alez Le\'on, L. Losano, J. Mateos Guilarte, and J. R. L. Santos, Construction of new scalar field models from the standard $\phi^4$ theory, {\it Phys. Scr.}  {\bf 87}  (2013),  045101.
\bibitem{bazeia_04} D. Bazeia, and A.R. Gomes, Bloch Brane, {\it JHEP}  {\bf 0405}  (2004), 012.
\bibitem{bazeia_042} D. Bazeia, C. Furtado, and A.R. Gomes, Brane Structure from a Scalar Field in Warped Spacetime, {\it JCAP}  {\bf 0402}  (2004), 002.
\bibitem{guth/1981} A. H. Guth, Inflationary universe: A possible solution to the horizon and flatness problems, {\it Phys. Rev. D}  {\bf 23}  (1981), 347.
\bibitem{ryden/2003} B. Ryden, {\it Introduction to Cosmology} (Addison Wesley, San Francisco, USA, 2003).
\bibitem{cognola/2009} G. Cognola, E. Elizalde, S.D. Odintsov, P. Tretyakov, and S. Zerbini., Initial and final de Sitter universes from modified $f(R)$ gravity, {\it Phys. Rev. D}  {\bf 79}  (2009), 044001.
\bibitem{namjoo/2012} M. Namjoo, H. Firouzjahi, and M. Sasaki, Multiple Inflationary Stages with Varying Equation of State, {\it JCAP}  {\bf 12}  (2012), 018.
\bibitem{ellis} J. Ellis, M. Fairbairn, and M. Sueiro, Rescuing Quadratic Inflation, {\it JCAP}  {\bf 02}  (2014), 044.
\bibitem{mukhanov} V. Mukhanov, {\it Physical Foundations of Cosmology} (Cambridge University Press, Cambridge UK, 2005).
\bibitem{planck15} P. A. R. Ade et al. [Planck Collaboration], Planck 2015 results XIII. Cosmological parameters, {\it A \& A}  {\bf 594} (2016), A13, astro-ph/1502.01589.
\bibitem{starobinsky}A. A. Starobinsky, A New Type of Isotropic Cosmological Models Without Singularity, {\it Phys. Lett. B}  {\bf 91}  (1980), 99.
\bibitem{aviles/2012} A. Aviles , C. Gruber, O. Luongo, and H. Quevedo, Cosmography and constraints on the equation of state of the Universe in various parametrizations, {\it Phys. Rev. D}  {\bf 86}  (2012), 123516.
\bibitem{gruber}C. Gruber, and O. Luongo, Cosmographic analysis of the equation of state of the universe through Pad\'e approximations, {\it Phys.Rev. D}  {\bf 89}  (2014), 103506.
\bibitem{almeida/2017} C. A. G. Almeida, M. A. Anacleto, F. A. Brito, E. Passos, and J. R. L. Santos, Cosmology in the Universe with Distance Dependent Lorentz-Violating Background, {\it Adv. High Energy Phys.}  {\bf 2017}  (2017), 5802352.
\bibitem{chand/2004} H. Chand, R. Srianand, P. Petitjean, and B. Aracil, Probing the cosmological variation of the fine-structure constant: Results based on VLT-UVES sample, {\it A \& A} {\bf 417}  (2004), 853.
\bibitem{nunes/2017} R. C. Nunes, A. Bonilla, S. Pan, and E. N. Saridakis, Observational constraints on $f(T)$ gravity from varying fundamental constants, {\it Eur. Phys. J. C}  {\bf 77}  (2017), 230.
\bibitem{mota/2004} D. F. Mota, and J. D. Barrow, Varying alpha in a more realistic universe, {\it Phys. Lett. B}  {\bf 581}  (2004), 141.
\bibitem{fritzsch/2017} H. Fritzsch, J. Sola, and R. C. Nunes, Running vacuum in the Universe and the time variation of the fundamental constants of Nature, {\it Eur. Phys. J. C}  {\bf 77}  (2017), 193.
\bibitem{uzan/2011} J. P. Uzan, Varying Constants, Gravitation and Cosmology, {\it Liv. Rev. Rel.}  {\bf 14}  (2011), 155.
\bibitem{szydlowski/2015} M. Szydlowski, Cosmological model with decaying vacuum energy from quantum mechanics, {\it Phys. Rev. D}  {\bf 91}  (2015), 123538.
\bibitem{bessada/2013} D. Bessada, and O.D. Miranda, Probing a cosmological model with a $\Lambda=\Lambda_0+3\beta H^{\,2}$ decaying-vacuum, {\it Phys. Rev. D}  {\bf 88}  (2013), 083530.
\bibitem{szydlowski/2017} M. Szydlowski, and A. Stachowski, Cosmological models with running cosmological term and decaying dark matter, {\it Phys. Dark Univ.}  {\bf 15}  (2017), 96.
\bibitem{Sonoda} T. Sonoda, Cosmological constant decaying with CMB temperature, {\it Int. J. Geom. Meth. Mod. Phys.}  {\bf 16}  (2019), 1950088.
\bibitem{motohashi} H. Motohashi, A. A. Starobinsky, and J. Yokoyama, Inflation with a constant rate of roll, {\it JCAP}  {\bf 09}  (2015), 018.
\bibitem{galvez} J. T. Galvez Ghersi, A. Zucca, and A. V. Frolov, Observational constraints on constant roll inflation, {\it JCAP}  {\bf 05}  (2019), 030.
\bibitem{bms_19}S. Basilakos, N. E. Mavromatos, and J. Sola, Scalar Field Theory Description of the Running Vacuum Model: from the scalaron to the ''vacuumon", gr-qc/1901.06638.
\bibitem{ml2002} J. M. F. Maia, and J. A. S. Lima, 
Scalar field description of decaying-$\Lambda$ cosmologies, {\it Phys. Rev. D}  {\bf 65}  (2002), 083513.

\end{thebibliography}
\end{document}